\documentclass[reqno]{amsart}
\usepackage{amsmath}%
\usepackage{amsfonts}%
\usepackage{amssymb}%
\usepackage{graphicx}
\usepackage{setspace}
\usepackage{microtype}

\theoremstyle{plain}

\numberwithin{equation}{section}
\begin{document}
\onehalfspacing
\raggedbottom
\title{Introducing Inner Nested Sampling}
\author{H.R.N.~van~Erp}

\author{R.O.~Linger}
\author{P.H.A.J.M. van Gelder}


\begin{abstract}
In this paper we will give a Monte Carlo algorithm by which the moments of a functions of Dirichlet probability distributions can be estimated. This algorithm is called Inner Nested Sampling and is an implementation of Skilling's general Nested Sampling framework. 
\end{abstract}

\maketitle

\section{Introduction}
This paper is for the benefit of those readers who are interested in finding moments of functions of Dirichlet probability distributions. In a data analysis we will, typically, have some set of count data, say,
\[		
	D = \left\{r_{1}, r_{2}, \ldots, r_{n}\right\},
\]
which by way of a Dirichlet distribution,   
\begin{equation}
	\label{eqInt4}
		p\!\left(\boldsymbol{\theta}\right) = \frac{\left(\sum r_{i} - 1\right)!}{\left(r_{1} - 1\right)! \left(r_{1} - 2\right)! \cdots \left(r_{n} - 1\right)!} \ \theta_{1}^{r_{1}-1} \theta_{2}^{r_{2}-1}\cdots \theta_{n}^{r_{n}-1},
\end{equation}
points to some set of underlying probabilities $\boldsymbol{\theta}$, where
\[
		\boldsymbol{\theta} = \left\{\theta_{1}, \theta_{2}, \ldots, \theta_{n}\right\}.
\]

Stated differently, the $\boldsymbol{\theta}$ are not directly observable, they can only be inferred. So, if we wish to assign a function $u$ to $\boldsymbol{\theta}$, then we have to take our uncertainty, in regards to the actual value of the vector $\boldsymbol{\theta}$, into account. But if we do so, then this will give us highly dimensional and highly intractable integrals. In order to evaluate these integrals, we will invoke the Nested Sampling \textit{framework}. 

In order to \textit{implement} this Nested Sampling framework, we will then proceed to introduce an algorithm, called Inner Nested Sampling. This algorithm is, for the specific problem of transforming the probability distribution $\boldsymbol{\theta}$ to a probability distribution of the function $u\!\left(\boldsymbol{\theta}\right)$, the optimal implementation, in that it takes into the specific geometry of the probability distribution \eqref{eqInt4}.

\section{Why Nested Sampling?}
In order to answer question why one ought to use Nested Sampling (NS) as their Monte Carlo framework of choice, we quote Skilling himself (personal communication, 2013): 
\begin{quote}
NS is \textit{designed} to get the evidence value, which is equation \#1 of inference.
\begin{quote}
		(1) \quad  Z = INTEGRAL dtheta JointProb(theta,data)\\
		(2) \quad  Posterior(theta) = JointProb(theta,data) / Z
\end{quote}
Logically, Z precedes the posterior and should never be ignored. It's the more important half of Bayes, not only in the algebra, but also because there's no point in bothering to generate a posterior if the evidence falls much short of some other plausible model. 
\end{quote} 
Stated differently, let $p\!\left(\theta,D\right)$ be the product of some prior $p\!\left(\theta\right)$ of some unknown set of parameters $\theta$ and some likelihood function $L\!\left(\theta, D\right)$ of some data set $D$:
\begin{equation}
	\label{eqInt1}
 	p\!\left(\theta,D\right) = p\!\left(\theta\right) L\!\left(\theta,D\right).
\end{equation}
Then the integral
\begin{equation}
	\label{eqInt2}
	Z = \int p\!\left(\theta,D\right) d\theta
\end{equation}
is the \textit{evidence} measure which may be used to differentiate between competing models, by way of Bayesian model selection. Furthermore, the posterior of the parameters $\theta$, given the data $D$, that is, $p\!\left(\left.\theta\right|D\right)$, is given as:
\begin{equation}
	\label{eqInt3}
	p\!\left(\left.\theta\right|D\right) =\frac{p\!\left(\theta,D\right)}{Z}.
\end{equation}

The Nested Sampling algorithm \cite{Skilling04, Skilling06} is specifically designed to evaluate the integral \eqref{eqInt2}, giving us an estimate of the evidence $Z$. Furthermore, it also provides us with a set of representative samples from the posterior \eqref{eqInt3}, which may function as a proxy for that posterior. 

For those cases where the integral \eqref{eqInt2} may be evaluated analytically, one will have no need for the Nested Sampling algorithm. However, for those problems where the integral \eqref{eqInt2} is both intractable and highly dimensional, there one will have to take his recourse to Nested Sampling, in order to be able to evaluate the evidence \eqref{eqInt2} and obtain a set of representative samples from the desired posterior \eqref{eqInt3}. 

The larger the evidence, the larger the adequacy of \eqref{eqInt1}, in terms of both parsimony and data fit. Stated differently, if we are considering some $p\!\left(\theta,D\right)$, then we only need to bother to construct a posterior for the model which has the largest evidence value. For what use are our (admittedly accurate) parameter estimates, as captured in the posterior distribution, if the model that leads us to these estimates is woefully inadequate? 

Before proceeding to the Nested Sampling algorithm, we will elaborate on the evidence measure and its role in Bayesian model selection. This is done for the benefit of those of the readers who are not yet well acquainted with these concepts. 

%
%
%

\section{The Evidence in Bayesian Model Selection}
Bayesian statistics has four fundamental constructs, namely, the prior, the likelihood, the posterior, and the evidence. These constructs are related in the following way:
\begin{equation}
	\label{eq1}
	\text{posterior} = \frac{\text{prior} \times \text{likelihood}}{\text{evidence}}.
\end{equation}
As an aside, any student of Bayesian probability theory will tend to have a firm grip on the concepts of a prior, likelihood, and posterior. However, the concept of evidence is less universally known. A possible explanation for this is that most people (used to) come to Bayesian probability theory by way of the more compact relationship
\begin{equation}
	\label{eq1b}
	\text{posterior} \propto \text{prior} \times \text{likelihood},
\end{equation}
which does not make any explicit mention of the evidence construct\footnote{See for example \cite{Zellner71} throughout.}. 

In what follows, we will employ in our analysis the correct, though notationally more cumbersome, relation \eqref{eq1}, and forgo of the more compact, but incomplete, Bayesian shorthand \eqref{eq1b}. This is done so the reader may develop some feeling for the evidence construct, and how this construct relates to the other three Bayesian constructs of prior, likelihood, and posterior. 

Let $p\left(\left.\theta\right|I\right)$ be the prior of some parameter $\theta$, where $I$ is the prior information regarding the unknown $\theta$ which we have to our disposal. Let $p\left(\left.D\right|\theta, M\right)$ be the probability of the data $D$ conditional on the value of parameter $\theta$ and some likelihood model $M$ which is used\footnote{Note that $M$ is a label that points to some likelihood model, and not so much a parameter whose value we wish to determine.}; the probability of the data is also known as the likelihood of the parameter $\theta$. Let $p\left(\left.\theta\right|D, M, I\right)$ be the posterior distribution of the parameter $\theta$, conditional on the data $D$, the likelihood model $M$, and the prior information model $I$. Then
\begin{equation}
	\label{eq2}
	p\left(\left.\theta\right|D, M, I\right) =  \frac{p\left(\left.\theta\right|I\right) p\left(\left.D\right|\theta, M\right)}{\int p\left(\left.\theta\right|I\right) p\left(\left.D\right|\theta, M\right) d\theta} =  \frac{p\left(\left.\theta\right|I\right) p\left(\left.D\right|\theta, M\right)}{ p\left(\left.D\right|M, I\right)},
\end{equation}
where
\begin{equation}
	\label{eq3}
p\left(\left.D\right|M, I\right) = \int p\left(\left.\theta, D\right|M, I\right) d\theta = \int p\left(\left.\theta\right|I\right) p\left(\left.D\right|\theta, M\right) d\theta
\end{equation}
is the evidence, that is, the marginalized likelihood, of both the likelihood model $M$ and the prior information model $I$. We now will show how this evidence may be used in Bayesian model selection.


If we have a set of likelihood models $M_{j}$ we wish to choose from, and just the one prior information model $I$, then we may do so by computing the evidence values $p\left(\left.D\right|M_{j}, I\right)$. 

Let $p\left(M_{j}\right)$ and $p\left(\left.M_{j}\right|D, I\right)$ be, respectively, the prior and posterior probability of the likelihood model $M_{j}$. Then the posterior probability distribution of these likelihood models is given as
\begin{equation}
	\label{eq4}
p\left(\left.M_{j}\right|D, I\right) = \frac{p\left(M_{j}\right) p\left(\left.D\right|M_{j}, I\right)}{\sum_{j} p\left(M_{j}\right). p\left(\left.D\right|M_{j}, I\right)}
\end{equation}
Note that if $p\left(M_{j}\right) = p\left(M_{k}\right)$ for  all $j$ and  $k$, then we have that \eqref{eq4} reduces to
\begin{equation}
	\label{eq5}
p\left(\left.M_{j}\right|D, I\right) = \frac{p\left(\left.D\right|M_{j}, I\right)}{\sum_{j} p\left(\left.D\right|M_{j}, I\right)}.
\end{equation}
Stated differently, if we assign equal prior probabilities to our different models, then these models may be ranked by their respective evidence values $p\left(\left.D\right|M_{j}, I\right)$, \cite{MacKay03}.

\section{Nested Sampling: The Idea}
By reducing any $k$-variate function $f$ to a corresponding monotonic descending univariate function $g$, and by using order statistics, the integral of any $k$-variate function $f$ may be evaluated using a Monte Carlo sampling scheme called Nested Sampling, \cite{Skilling06}.

\begin{figure}[!h]
		\centering
			\includegraphics[width=0.60\textwidth]{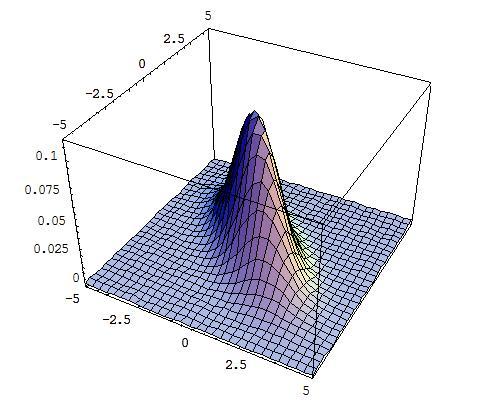}
		\caption{Plot of Function $f$}
		\label{plot6a}
	\end{figure}

\begin{figure}[!h]
	\centering
		\includegraphics[width=0.60\textwidth]{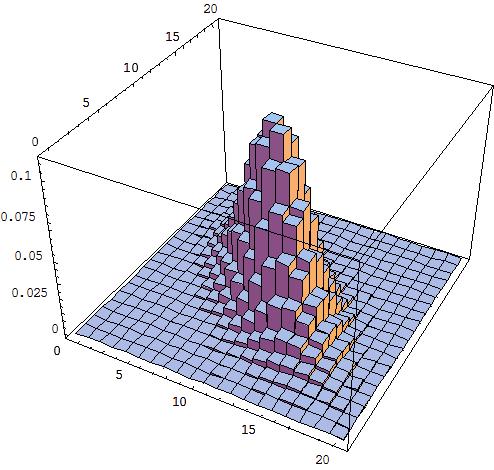}
	\caption{Volume Elements of Function $f$}
	\label{pictureChap6b}
\end{figure}

Say, we wish to numerically evaluate a bivariate distribution $f\!\left(x,y\right)$, where
\begin{equation}
	\label{eq.6.1}
	f\!\left(x,y\right) = \frac{\sqrt{1-0.7^2}}{2\pi} \exp\left[-\frac{1}{2}\left(x^{2}+1.4 x y + y^{2}\right) \right]
\end{equation}
where $-5 \leq x, y\leq 5$, Figure~\ref{plot6a}.
The total volume under the curve \eqref{eq.6.1} is given by the integral 
\begin{equation}
	\label{eq.6.2}
	\int^{5}_{-5}\int^{5}_{-5} \frac{\sqrt{1-0.7^2}}{2\pi} \exp\left[-\frac{1}{2}\left(x^{2}+1.4 x y + y^{2}\right) \right] dx\; dy = 0.9993
\end{equation}
We may evaluate the integral \eqref{eq.6.1} through brute computational force. Say, we partition the  $x,y$-plane in little squares with areas $dx_{j} dy_{k} = 0.25$, for $j=1,\ldots,20$ and $k=1,\ldots,20$. Then define the center of these areas as $\left(\tilde{x}_{j},\tilde{y}_{k}\right)$, and compute the strips of volume $V_{j k}$ as 
 \begin{equation}
	\label{eq.6.3}
	V_{j k} = f\!\left(\tilde{x}_{j},\tilde{y}_{k}\right) dx_{j} dy_{k}
\end{equation}
In Figure~\ref{pictureChap6b} we give all the volume elements $V_{j k}$  together.

\noindent Summating the volume elements \eqref{eq.6.3}, the total volume under the curve \eqref{eq.6.1} may be approximated as
\begin{equation}
	\label{eq.6.4}
	\text{volume} \approx \sum^{20}_{j = 1}\sum^{20}_{k = 1} V_{j k} = 0.9994
\end{equation}

\begin{figure}[!h]
	\centering
		\centerline{\includegraphics[width=0.60\textwidth]{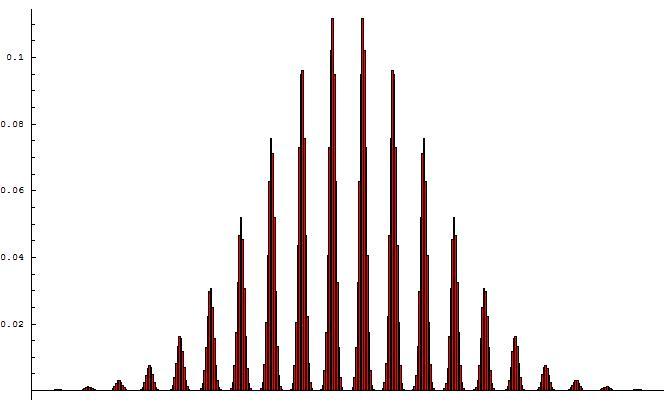}}
	\caption{Area Elements of Function $f$}
	\label{pictureChap6c}
\end{figure}

The 3-dimensional volume elements $V_{j k}$  may be mapped to corresponding 2-dimensional area elements $A_{i}$, without any loss of generality. In order to demonstrate this, we introduce the following notation
\begin{equation}
	\label{eq.6.5}
	dw_{i} = dx_{j} \: dy_{k}, \quad f_{i} = f\!\left(\tilde{x}_{j},\tilde{y}_{k}\right) 
\end{equation}
where index $i$ is a function of the indices $j$ and $k$:
\begin{equation}
	\label{eq.6.5b}
	i \equiv \left(j-1\right)20 + k
\end{equation}
and $i = 1, \ldots, 400$. Using \eqref{eq.6.5}, we may rewrite \eqref{eq.6.3} as
\begin{equation}
	\label{eq.6.6}
	V_{j k} = f\!\left(\tilde{x}_{j},\tilde{y}_{k}\right) dx_{j} dy_{k} = f_{i} \: dw_{i} = A_{i} 
\end{equation}
In Figure~\ref{pictureChap6c} we give all the 400 area elements $A_{i}$ together.

\begin{figure}[!h]
	\centering
		\centerline{\includegraphics[width=0.60\textwidth]{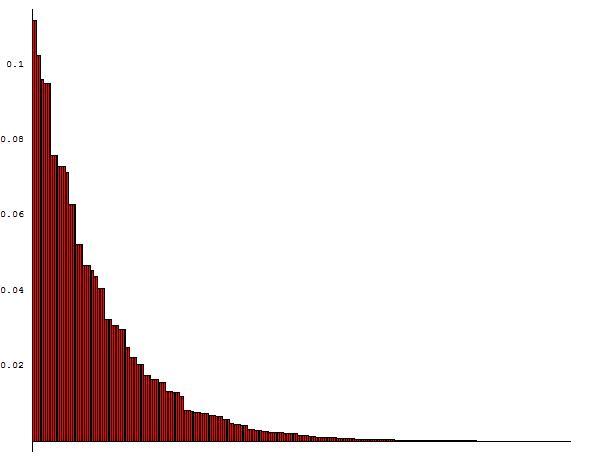}}
	\caption{Ordered Area Elements of Function  $f$}
	\label{pictureChap6d}
\end{figure}

Since \eqref{eq.6.6} is equivalent to \eqref{eq.6.3}, we have that the mapping of the 3-dimensional volume elements, $V_{j k}$,  to their corresponding 2-dimensional area elements, $A_{i}$,  has not led to any loss of information; that is,
\begin{equation}
	\text{area} = \sum^{400}_{i = 1} A_{i} =  \sum^{20}_{j = 1}\sum^{20}_{k = 1} V_{j k} =\text{volume} 
\end{equation}
We may, trivially, rearrange the elements $A_{i}$ in Figure~\ref{pictureChap6c} in descending order, Figure~\ref{pictureChap6d}.

\begin{figure}[!h]
	\centering
		\centerline{\includegraphics[width=0.60\textwidth]{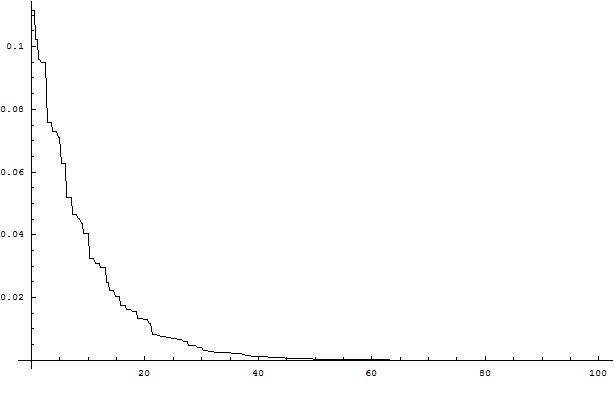}}
	\caption{Plot of Hypothetical Function $g$}
	\label{pictureChap6e}
\end{figure}

\noindent All these rectangular elements have a base of  $dw = dx\:dy = 0.25$. So, being that there are 400 area elements, we might view Figure~\ref{pictureChap6d} as a representation of some hypothetical monotonic descending function $g\!\left(w\right)$, where  $0\leq w\leq 100$, Figure~\ref{pictureChap6e}.

What we have accomplished is that we have mapped 3-dimensional volume elements, Figure~\ref{pictureChap6b}, to 2-dimensional area elements, Figure~\ref{pictureChap6c}, after which we have rearranged the area elements, Figure~\ref{pictureChap6d}, so as to get a monotonic descending function $g\left(w\right)$,  Figure~\ref{pictureChap6e}. If we integrate the univariate function $g\left(w\right)$, Figure~\ref{pictureChap6e}, we get the volume we are looking for. 

Any $k$-variate function may be reduced to its corresponding monotonic descending univariate function $g\!\left(w\right)$. We will see that the procedure of Nested Sampling is based upon the equivalence between any $k$-variate function and its corresponding univariate representation $g\!\left(w\right)$.

\subsection{Estimating abscissa's} 
Skilling's Nested Sampling framework is unique in that it is a Monte Carlo scheme that uses probability theory to its advantage, \cite{Skilling06}. 
 
If we have a value of  $g\!\left(w\right)$, without knowing the value of $w$. Then the only thing we know about $w$ is that it must lie somewhere in the region $0\leq w\leq W$, where $W$ is the scalar value of the volume on which the $k$-variate function is defined. 

So, if, based on this information alone, we were asked to assign a probability distribution to the actual value of the abscissa $w$ that corresponds with the ordinate $g\!\left(w\right)$, then we would have to assign a univariately uniformly distribution on the domain $\left[0,W\right]$, that is, 
\begin{equation}
	\label{eq.6.7bq}
	p\!\left(w\right) = \frac{1}{W}, \qquad 0 \leq w \leq W. 
\end{equation}
Consequently, the unknown abscissa value $w$ has an expectation value of
\begin{equation}
	\label{eq.6.7}
	E\!\left(w\right) = \frac{W}{2}
\end{equation}
and a standard deviation of
\begin{equation}
	\label{eq.6.7q}
	\text{std}\!\left(w\right) = \frac{W}{2\sqrt{3}}.
\end{equation}

Let $\boldsymbol{x} = \left(x_{1}, \ldots, x_{k}\right)$ be a point in the $k$-dimensional domain. If we have $N$ random samples of the $k$-variate domain, that is, $\left\{\boldsymbol{x}_{1},\ldots,\boldsymbol{x}_{N}\right\}$. Then, by evaluating these points, we have a set of $N$ random ordinates, that is, $\left\{f\!\left(\boldsymbol{x}_{1}\right),\ldots,f\!\left(\boldsymbol{x}_{N}\right)\right\}$. Because of the equivalence between any $k$-variate function $f$ and its univariate representation $g$, we may write
\begin{equation}
	\label{eq.6.7b}
	g\!\left(w_{n}\right) = f\!\left(\boldsymbol{x}_{n}\right), \qquad  n = 1, \ldots, N
\end{equation}
So, by construction, the set of $N$ random ordinates $\left\{f\!\left(\boldsymbol{x}_{1}\right),\ldots,f\!\left(\boldsymbol{x}_{N}\right)\right\}$ corresponds with the set of random ordinates $\left\{g\!\left(w_{1}\right), \ldots , g\!\left(w_{N}\right)\right\}$. 

Even though we have the ordinate $g\!\left(w_{n}\right)$, by way of \eqref{eq.6.7b}, we do not know the abscissa $w_{n}$ that goes with this ordinate. As we sample the $k$-variate domain $\boldsymbol{x}$, rather than the univariate domain $w$. However, the one thing we do know is that the smallest realisation of $g\left(w_{n}\right)$ corresponds with the greatest value of $w_{n}$. This is because function $g$ is a monotonic descending function, see Figure~\ref{pictureChap6e}. It follows that we may use an order distribution for the unknown value $\max\left(w_{n}\right)$, or, for short, $w_{\max}$. 

The order probability distribution for the largest draw, $w_{\max}$, in a sample of size $N$ from the uniform distribution \eqref{eq.6.7b} is
\begin{equation}
	\label{eq.6.8}
	p\!\left(w_{\max}\right) = N \left(\frac{w_{\max}}{W}\right)^{N-1} \frac{1}{W}
\end{equation}
where both $N$ and $W$ are known. The order distribution \eqref{eq.6.8} has a mean of
\begin{equation}
	\label{eq.6.9}
	E\!\left(w_{\max}\right) = \left(1-\frac{1}{N+1}\right) W
\end{equation}
and a standard deviation of
\begin{equation}
	\label{eq.6.9b}
 \text{std}\!\left(w_{\max}\right) = \sqrt{\frac{N}{\left(N+1\right)^{2}\left(N+2\right)}} W
\end{equation}
We see in \eqref{eq.6.9b} that the standard deviation, that is, our uncertainty regarding the unknown value of $w_{\max}$ falls of with a factor $N$. We will see that \eqref{eq.6.9} and \eqref{eq.6.9b} form the backbone of the Nested Sampling framework. 

Note that Skilling himself does not arrive at Nested Sampling algorithm by way of order statistics, but by way of the closely related probability distribution of the shrinkage ratio of the admissible domain of the unknown univariate representation $g$ of the multivariate function $f$: 
\begin{equation}
	\label{eq.6.9d}
	t = \frac{w_{\max}}{W}.
\end{equation}
This shrinkage ratio has a probability distribution 
\begin{equation}
	\label{eq.6.9e}
	p\left(t\right) = N t^{N-1}.
\end{equation}
And we may compare \eqref{eq.6.8} with \eqref{eq.6.9e}, to see that both probability distributions are equal, up to a change of variable, which scales the maximum abscissa value $w_{\max}$ to the fraction $t$, \eqref{eq.6.9d}.


%


\subsection{Nested Sampling: The Algorithm}
In this initial version of the Nested Sampling framework we will not protect against under- and overflow. We will just focus here on the basic philosophy which underlies Nested Sampling. 

In what follows we reserve subscripts for sample membership and superscripts for the enumeration of the algorithmic step. 
\\
\\
\noindent \textbf{Step 1}
\\
\noindent Find $N$ random values $\boldsymbol{x}$ in the $k$-variate domain and evaluate, that is, find
\begin{equation}
	\label{eq.6.9c}
	f\!\left(\boldsymbol{x}_{n}^{\left(1\right)}\right), \quad \quad  n = 1, \ldots , N
\end{equation}
Since we may perform the steps as shown in Figure~\ref{plot6a} through Figure~\ref{pictureChap6e}, it holds, by way of construction, that the $N$ values of $f\!\left(\boldsymbol{x}_{n}^{\left(1\right)}\right)$ correspond with $N$ values of $g\!\left(w_{n}\right)$, \eqref{eq.6.7b}. 

As stated before, we cannot explicitly map the abscissa $\boldsymbol{x}$ to the corresponding abscissa of the monotonic descending function, $w$. We could link the volume elements \eqref{eq.6.3} to the area elements\eqref{eq.6.7} using brute computational force. But this would amount to regular numerical integration and, thus, we would still be faced with the curse of dimensionality. So, instead, we use \eqref{eq.6.9} to statistically approximate the $w$ which corresponds with the smallest sampled value of $f\!\left(\boldsymbol{x}_{n}^{\left(1\right)}\right)$ and, thus, get our first coordinate $\left[w^{\left(1\right)}, g^{\left(1\right)}\right]$ of the unknown function $g$, Figure~\ref{pictureChap6e}. The abscissa and ordinate of this coordinate are, respectively,
\begin{equation}
	\label{Alg1}
	w^{\left(1\right)} =  \left(1-\frac{1}{N+1}\right) W
\end{equation}
and
\begin{equation}
	\label{Alg1b}
	g^{\left(1\right)} = \min f\!\left(\boldsymbol{x}_{n}^{\left(1\right)}\right)
\end{equation}
As the error of the statistically approximated $w^{\left(1\right)}$ falls of with a factor $N$, \eqref{eq.6.9b}, the accuracy of the approximation \eqref{Alg1} will increase as $N$, the number of random samples, increases. 

Finally, we approximate the integral to the right of $w^{\left(1\right)}$ of our constructed function $g$, Figure~\ref{pictureChap6e}, as
\begin{equation}
	\label{Alg2}
	A^{\left(1\right)} = \left(W - w^{\left(1\right)}\right) g^{\left(1\right)} = \frac{W}{N+1} g^{\left(1\right)} \approx  \int^{W}_{w_{1}} g\!\left(w\right) dw
\end{equation}
and set the evaluated integral at the first step, $Z^{\left(1\right)}$, to
\begin{equation}
	\label{Alg3}
	Z^{\left(1\right)}= A^{\left(1\right)}
\end{equation}      
\\
\\
\\
\noindent \textbf{Step 2}
\\
\noindent We again find $N$ random values $\boldsymbol{x}$ in the $k$-variate domain and evaluate. But now we constrain the random values $\boldsymbol{x}$ in that their mapping to the function $f$ be equal or greater than the minimum value of the mapping of the previous mapping, that is, we sample the $\boldsymbol{x}_{n}^{\left(2\right)}$ under the constraint
\begin{equation}
	\label{Alg4}
	f\!\left(\boldsymbol{x}_{n}^{\left(2\right)}\right) \geq \min f\!\left(\boldsymbol{x}_{m}^{\left(1\right)}\right),			\quad \quad n, m = 1, \dots, N
\end{equation}  

In the first iteration our state of knowledge was that $0 \leq w \leq W$. In the second iteration we sample $\boldsymbol{x}_{n}^{\left(2\right)}$ under constraint \eqref{Alg4}, which is equivalent to sampling $w$ under the constraint, \eqref{Alg1b},
\begin{equation}
	\label{Alg5}
	g\left(w\right) \geq g^{\left(1\right)}	
\end{equation}  
where $g$ is a monotonic descending function in $w$. So, the updated state of knowledge is that $0 \leq w \leq w^{\left(1\right)}$.

Using \eqref{eq.6.9} again, but replacing $W$, the initial upperbound of $w$, with $w^{\left(1\right)}$, the new upperbound of $w$, we may approximate the second coordinate $\left[w^{\left(2\right)}, g^{\left(2\right)}\right]$ of the unknown function $g$:
\begin{equation}
	\label{Alg6}
	w^{\left(2\right)} =  \left(1-\frac{1}{N+1}\right) w^{\left(1\right)} = \left(1-\frac{1}{N+1}\right)^{2} W
\end{equation}
and
\begin{equation}
	\label{Alg6b}
	g^{\left(2\right)} = \min f\!\left(\boldsymbol{x}_{n}^{\left(2\right)}\right) 
\end{equation}

We approximate the area of the integral between $w^{\left(2\right)}$ and $w^{\left(2\right)}$, where $w^{\left(2\right)} < w^{\left(1\right)}$, as
\begin{equation}
	\label{Alg7}
	A^{\left(2\right)} = \left(w^{\left(1\right)} - w^{\left(2\right)}\right) g^{\left(2\right)} = \frac{w^{\left(1\right)}}{N+1} g^{\left(2\right)} \approx \int^{w^{\left(1\right)}}_{w^{\left(2\right)}} g\left(w\right) dw 
\end{equation}
and set the evaluated integral at the second step, $Z^{\left(2\right)}$, to
\begin{equation}
	\label{Alg8}
	Z^{\left(2\right)}= A^{\left(1\right)}  + A^{\left(2\right)} 
\end{equation}   
\\
\\
\noindent \textbf{Step \textit{t}}
\\
\noindent In iteration step $t$ we approximate coordinate $\left[w^{\left(t\right)}, g^{\left(t\right)}\right]$ of the unknown function $g$ by way of
\begin{equation}
	\label{Alg9}
	w^{\left(t\right)} =  \left(1-\frac{1}{N+1}\right) w^{\left(t-1\right)} = \left(1-\frac{1}{N+1}\right)^{t} W
\end{equation}
and
\begin{equation}
	\label{Alg9b}
	g^{\left(t\right)} = \min f\!\left(\boldsymbol{x}_{n}^{\left(t\right)}\right)
\end{equation}
where the $\boldsymbol{x}_{n}^{\left(t\right)}$ have been sampled under the constraint
\begin{equation}
	\label{Alg9c}
	f\!\left(\boldsymbol{x}_{n}^{\left(t\right)}\right) \geq \min f\!\left(\boldsymbol{x}_{m}^{\left(t-1\right)}\right),			\quad \quad n, m = 1, \dots, N
\end{equation}  

The area of the integral between $w^{\left(t\right)}$ and $w^{\left(t-1\right)}$, where $w^{\left(t\right)} < w^{\left(t-1\right)}$, is approximated as
\begin{equation}
	\label{Alg10}
	A^{\left(t\right)} = \left(w^{\left(t-1\right)} - w^{\left(t\right)}\right) g^{\left(t\right)} =  \frac{ w^{\left(t-1\right)}}{N+1} g^{\left(t\right)}  \approx \int^{w^{\left(t-1\right)}}_{w^{\left(t\right)}} g\left(w\right) dw 
\end{equation}
The evaluated integral at the step $t$, $Z^{\left(t\right)}$, is updated as
\begin{equation}
	\label{Alg11}
	Z^{\left(t\right)}= Z^{\left(t-1\right)} + A^{\left(t\right)}  = \sum^{t}_{i=1} A^{\left(i\right)} 
\end{equation}  
\\
\\
\noindent \textbf{Termination step}
\\
\noindent Because of the identity \eqref{Alg9}
\[
	w^{\left(t\right)} = \left(1-\frac{1}{N+1}\right)^{t} W
\]
we have that $w^{\left(t\right)} \rightarrow 0$ as $t \rightarrow \infty$. If $f$ is bounded, then by construction so is $g$. It then follows from \eqref{Alg10} that bounded functions $f$ or, equivalently, bounded $g$
\begin{equation}
\label{Alg11b}
	A^{\left(t+1\right)} = \frac{w^{\left(t\right)}}{N+1} g^{\left(t+1\right)} \rightarrow 0
\end{equation}
as $t \rightarrow \infty$. The maximum of $f$, the function whose integral we wish to evaluate, is also the bound of $g$, its univariate representation. This implies the following inequality 
\begin{equation}
	\label{Alg12a}
	g^{\left(t+1\right)}  \leq  \max f\!\left(\boldsymbol{x}\right)
\end{equation}  
By way of \eqref{Alg10} and \eqref{Alg12a}, we then have that
\begin{equation}
	\label{Alg12b}
	A^{\left(t+1\right)} \leq \frac{w^{\left(t\right)}}{N+1} \max f\!\left(\boldsymbol{x}\right)
\end{equation}
So, if we can determine the maximum of $f$, we can use as a possible terminating condition the point where the upperbound of $A^{\left(t+1\right)}$ does not contribute more than $N^{2}$ part to $Z^{\left(t+1\right)}$
\begin{equation}
	\label{Alg13}
	\frac{w_{t}}{N+1} \max f\!\left(\boldsymbol{x}\right) < \frac{Z^{\left(t+1\right)}}{N^{2}}
\end{equation} 
Alternatively, if we are unable determine the maximum of $f$, then we take as a stopping criterium the point where $A^{\left(t+1\right)}$ itself does not contribute more than $N^{2}$ part to $Z^{\left(t+1\right)}$
\begin{equation}
	\label{Alg14}
	A^{\left(t+1\right)} < \frac{Z^{\left(t+1\right)}}{N^{2}}
\end{equation}  
However, \eqref{Alg14} comes with the caveat that the function $f$, or, equivalently, $g$, might still have some unexplored regions having large enough values $w^{\left(q+1\right)}$, where $q > t$, to take 
\begin{equation}
	\label{Alg14b}
A^{\left(q+1\right)}  = \frac{w^{\left(q\right)}}{N+1} g^{\left(q+1\right)} > \frac{Z^{\left(q+1\right)}}{N^{2}}
\end{equation}  
which would imply a premature termination of the algorithm.

\subsection{Optimal implementation}
If one is to implement the steps in the previous section one would get an algorithm very differently from the pseudo-code given by Skilling, \cite{Skilling06}. These differences are mainly differences in implementation. 

The reason for us to first give the naive algorithm was to point to reader to the elegant and simple idea behind Nested Sampling, without losing ourselves too much in the technicalities of optimal implementations. 

However, with the core idea behind Nested Sampling demonstrated, we now will treat in the next two paragraphs the points of optimal implementation.

\subsubsection{Reduction of computational cost}
\label{section6.4.1}
In the previous treatment of the Nested Sampling framework, $N$ new samples of $\boldsymbol{x}$  were drawn at each iteration step $t$, under the constraint
\begin{equation}
	\label{Alg15}
	f\!\left(\boldsymbol{x}_{n}^{\left(t\right)}\right) \geq \min f\!\left(\boldsymbol{x}_{m}^{\left(t-1\right)}\right),			\quad \quad n, m = 1, \dots, N
\end{equation}  
Now, instead of drawing $N$ new samples at each sampling step $t$, we may also realize that in iteration $\left(t-1\right)$, we already had $\left(N-1\right)$ samples of $f$ at our disposal that satisified the constraint \eqref{Alg15}. Namely, those samples in iteration step $\left(t-1\right)$ smaller than 
\[
	\min f\!\left(\boldsymbol{x}_{m}^{\left(t-1\right)}\right). 
\]
If we keep these $\left(N-1\right)$ samples, then we only need to sample one aditional ordinate value which satisfies the constraint \eqref{Alg15}, in order to obtain our needed sample of $N$ objects. 

So, after each iteration $t$ we discard one object from our sample of $N$ objects. This discarded object becomes $g^{\left(t\right)}$. The $\left(N-1\right)$ surviving objects are taken to the next iteration and an additional object is sampled under constraint
\[
	 f\!\left(\boldsymbol{x}_{n}^{\left(t+1\right)}\right) \leq g^{\left(t\right)}.
\] 
This implementation reduces the computational costs of Nested Sampling with an order of magnitude of $N$.

\subsubsection{Guarding against under- and overflow}
In many problems $Z^{\left(t\right)}$  may become so large that computational overflow may occur, that is, there is no longer a floating number representation possible for its value. To remedy this problem we will have to work with the $\log Z^{\left(t\right)}$. Furthermore, We also have that certain functions $f$ have certain values in their domain so small that computational underflow may occur. So, in what follows we will evaluate $\log f$  instead of $f$. Likewise, because of the fact that $w_t \rightarrow 0$ as $t \rightarrow \infty$, we have that for sufficiently large $t$ computational underflow may occur, that is, $w_{t}$ may become so small that there is no longer a floating number representation possible for its value. To remedy the latter situation we will go from the $w$ scale to the $\log w$ scale.

To go to the $u = \log w$ scale, we will have to make a proper change of variable for the order distribution \eqref{eq.6.8}. We have that 
\begin{equation}
	\label{eq.6.10}
	du = \frac{dw}{w}, \quad w = \exp u
\end{equation}
Substituting \eqref{eq.6.10} into \eqref{eq.6.8}, we may obtain
\begin{equation}
	\label{eq.6.11}
	p\!\left(u_{\max}\right) = \frac{N}{W^{N}} \exp\left( N u_{\max}\right)
\end{equation}
for $-\infty < u_{\max}\leq \log W$.
with mean standard deviation of
\begin{equation}
	\label{eq.6.12}
	E\!\left(u_{\max}\right) = -\frac{1}{N} + \log W
\end{equation}
and
\begin{equation}
	\label{eq.6.12b}
	\text{std}\!\left(u_{\max}\right) = \frac{1}{N}
\end{equation}
With a repeated application of \eqref{eq.6.12} we may find the limits of the $u$ scale after the $t$th iteration to be
\begin{equation}
	\label{eq.6.13}
	u^{\left(t\right)} = -\frac{t}{N} + \log W
\end{equation}
From \eqref{eq.6.13}, it follows that that the width of the $t$th interval on the original $w$ scale may be written as
\begin{align}
	\label{eq.6.14}
	w^{\left(t-1\right)} - w^{\left(t\right)} &= \exp\left(u^{\left(t-1\right)}\right) - \exp\left(u^{\left(t\right)}\right) \nonumber \\
	\nonumber \\
	&= \exp\left(-\frac{t-1}{N} + \log W\right) - \exp\left(-\frac{t}{N} + \log W\right) \nonumber \\
	 \\
	&= W\left[\exp\left(-\frac{t-1}{N} \right) - \exp\left(-\frac{t}{N} \right)\right] \nonumber \\
	\nonumber \\
	&= W\exp\left(-\frac{t-1}{N} \right)  \left[1 - \exp\left(-\frac{1}{N} \right)\right] \nonumber
\end{align}
Consequently, we have
\begin{equation}
	\label{eq.6.15}
	\log\left(w^{\left(t-1\right)} - w^{\left(t\right)}\right) = -\frac{t-1}{N} + \log W +\log\left[1 - \exp\left(-\frac{1}{N} \right)\right]
\end{equation}
We then have that that the log of the area element at the $t$th iteration is the sum
\begin{align}
	\label{eq.6.16}
	\log A^{\left(t\right)}  &= \log\left(w^{\left(t-1\right)} - w^{\left(t\right)}\right) + \log g^{\left(t\right)} \nonumber \\
	\\
	&= -\frac{t-1}{N} + \log W +\log\left[1 - \exp\left(-\frac{1}{N} \right)\right] + \log g^{\left(t\right)} \nonumber
\end{align}
With \eqref{eq.6.16} the computation of $A^{\left(t\right)}$ is sufficiently protected from underflow of both $w^{\left(t\right)}$ and $f$, or, equivalently, $g$.

We may update $\log Z^{\left(t\right)}$ with $\log A^{\left(t\right)}$ using the formula for logarithmic addition, (Skilling, 2006), 
\begin{align}
	\label{eq.6.17}
	\log \left(\exp x + \exp y\right) &= \log\left\{\exp x \left[1 + \exp \left(y-x\right) \right]\right\} \nonumber \\
	\\
	&= x + \log\left[1 + \exp \left(y-x\right) \right]\nonumber
\end{align}
If we set $x = \log Z^{\left(t-1\right)}$ and $y =\log A^{\left(t\right)}$, then \eqref{eq.6.17} gives us
\begin{equation}
	\label{eq.6.18}
	 \log Z^{\left(t\right)} = \log Z^{\left(t-1\right)}+ \log\left[1 + \exp \left(\log A^{\left(t\right)}-\log Z^{\left(t-1\right)}\right) \right]
\end{equation}
With \eqref{eq.6.18} the computation of $\log Z^{\left(t\right)}$ is sufficiently protected from the overflow of $\log Z^{\left(t-1\right)}$.

To summarize, in order to protect the algorithm from under- and overflow, which may easily occur in actual problems, the original algorithm remains unchanged, except that each iteration $\log A^{\left(t\right)}$ is computed, instead of $A^{\left(t\right)}$, by way of \eqref{eq.6.16}, and $\log Z^{\left(t\right)}$ is updated, instead of $Z^{\left(t\right)}$, by \eqref{eq.6.18}. Note that the termination conditions \eqref{Alg13} and  \eqref{Alg14} transform, respectively, to 
\begin{equation}
	\label{eq.6.19}
	-\frac{t}{N} + \log W +\log\left[1 - \exp\left(-\frac{1}{N}\right)\right] +  \log\left[\max f\!\left(\boldsymbol{x}\right)\right] < \log Z^{\left(t+1\right)} - 2\log N
\end{equation} 
if we know the maximum of the function $f$, and
\begin{equation}
	\label{eq.6.20}
	 \log A^{\left(t\right)} < \log Z^{\left(t+1\right)} - 2\log N
\end{equation}  
if we do not.

\section{Applying Nested Sampling to Unscaled Posteriors}
Nested Sampling is a general purpose algorithm for the evaluation of the integrals of multivariate functions. A special class of multivariate functions are unscaled posterior distributions. 

Let $p\!\left(x,y\right)$ be some prior distribution of the unknown parameters $x$ and $y$. Let $p\!\left(\left.D\right|x,y\right)$ be the probability of some observed data set $D$ given the parameters $x$ and $y$, or, equivalently, the likelihood $L\!\left(x,y\right)$. Then the unscaled posterior distribution is
\begin{equation}
	\label{eq.6.21}
  p\!\left(x,y,D\right) = p\!\left(x,y\right) p\!\left(\left.D\right|x,y\right) = p\!\left(x,y\right) L\!\left(x,y\right)
\end{equation}
If we integrate $p\!\left(x,y,D\right)$ over the domain of  $x$ and $y$ we get the marginal probability of the observed data set $D$
\begin{equation}
	\label{eq.6.22}
  p\!\left(D\right) = \int \int p\!\left(x,y,D\right) dx\;dy = \int \int  p\!\left(x,y\right) p\!\left(\left.D\right|x,y\right) dx\;dy 
\end{equation}
The marginal probility $p\!\left(D\right)$ is also called the \textit{evidence}. Scaling $p\!\left(x,y,D\right)$ with $p\!\left(D\right)$ we get, by way of the product rule, the scaled posterior distribution:
\begin{equation}
	\label{eq.6.23}
  p\!\left(\left.x,y\right|D\right) = \frac{p\!\left(x,y,D\right)}{p\!\left(D\right)}
\end{equation}
Furthermore, the evidence is an important quantity in its own right. It may be used for model selection. Say, we have two competing likelihood models, $L_{1}$ and $L_{2}$. Combining \eqref{eq.6.21} and \eqref{eq.6.22}, we may compete the respective evidences of the competing likelihood models:
\begin{align}
	\label{eq.6.24}
  p_{1}\!\left(D\right) &= \int \int p\!\left(x,y\right) L_{1}\!\left(x,y\right) dx\;dy \nonumber \\
  \\
  p_{2}\!\left(D\right) &= \int \int p\!\left(x,y\right) L_{2}\!\left(x,y\right) dx\;dy \nonumber
\end{align}
Then the likelihood model that fits the data best will be the model that gives us the greatest marginal probability of obtaining the observed data set $D$. The evaluation of the integral \eqref{eq.6.21} is a non-trivial matter, especially for highly variate likelihood functions. It was with this application in mind that Skilling developped his Nested Sampling framework.

We now proceed to apply the Nested Sampling framework to \eqref{eq.6.21}. Let the function $f$ be
\begin{equation}
\label{eq.6.25}
	f\!\left(x,y\right) = p\!\left(x,y\right) L\!\left(x,y\right)
\end{equation}
Let the prior $p\!\left(x,y\right)$ be the uniform distribution over the domain of $x$ and $y$. If this domain has a total area of $W$, then we have that
\begin{equation}
\label{eq.6.26}
	p\!\left(x,y\right) = \frac{1}{W}
\end{equation}
Substituting \eqref{eq.6.26} into \eqref{eq.6.25}, and substituting the resulting function $f$ into \eqref{eq.6.16}, we get
\begin{align}
	\label{eq.6.27}
	\log A^{\left(t\right)}  &= -\frac{t-1}{N} + \log W +\log\left[1 - \exp\left(-\frac{1}{N} \right)\right] + \log g^{\left(t\right)}  \nonumber \\
	\nonumber \\
	&= -\frac{t-1}{N} + \log W +\log\left[1 - \exp\left(-\frac{1}{N} \right)\right] -\log W + \log\left(\min L_{n}^{\left(t\right)}\right) \nonumber \\
	\\
	&= -\frac{t-1}{N} +\log\left[1 - \exp\left(-\frac{1}{N} \right)\right] + \log\left(\min L_{n}^{\left(t\right)}\right) \nonumber
\end{align}
where $\min L_{n}^{\left(t\right)}$ is the smallest likelihood from the sample of $N$ likelihoods at the the $t$th step, which were sampled under the constraint that
\begin{equation}
	\label{eq.6.28}
	L_{n}^{\left(t\right)} \leq \min L_{m}^{\left(t-1\right)}
\end{equation}

The Nested Sampling framework is most often presented in the form of \eqref{eq.6.27}, which is the special case of $f$ being an unscaled posterior wich takes as its prior the uniform distribution of the parameter space. However, the framework is much more general in that is applicable to any function $f$, not just unscaled posterior distributions. 

Last but not least, if $f$ is an unscaled posterior, then the Nested Sampling framework not only evaluates the evidence. It also gives us a Monte Carlo proxy of the posterior of interest. Let $T$ be the termination step of the Nested Sampling run. Let $\boldsymbol{x}^{\left(t\right)}$ be the points in the $k$-variate domain that corresponded with the $\boldsymbol{x}_{n}^{\left(t\right)}$ for which
\begin{equation}
	\label{eq.6.29}
	 g^{\left(t\right)} = \min  p\!\left(\boldsymbol{x}_{n}^{\left(t\right)}\right) L\!\left(\boldsymbol{x}_{n}^{\left(t\right)}\right)
\end{equation}
where $t = 1, \ldots, T$ and $n = 1, \ldots, N$. Then we may assign to these $\boldsymbol{x}^{\left(t\right)}$  the probability weights, \eqref{Alg10} and \eqref{Alg11},
\begin{equation}
	\label{eq.6.30}
	  P\!\left(\boldsymbol{x}^{\left(t\right)}\right) = \frac{A^{\left(t\right)}}{Z^{\left(T\right)}}
\end{equation}
This leaves us with a set of weighted random Monte Carlo samples. The weighted samples are a proxy for the posterior distribution of interest, \cite{Skilling06} . 

Let the posterior $p\!\left(\left.\boldsymbol{x}\right|D \right)$ be given as
\begin{equation}
	\label{eq.6.31} 
p\!\left(\left.\boldsymbol{x}\right|D \right) = \frac{p\!\left(\boldsymbol{x}\right) L\!\left(\boldsymbol{x}\right)}{\int p\!\left(\boldsymbol{x}\right) L\!\left(\boldsymbol{x}\right) d\boldsymbol{x}} 
\end{equation}
Let $h$ be some function defined on the $k$-variate domain of the $\boldsymbol{x}$. Then we may approximate the weighted $h$ by way of Nested Sampling as, \eqref{eq.6.30}, 
\begin{equation}
	\label{eq.6.32} 
\int h\!\left(.\boldsymbol{x}\right) p\!\left(\left.\boldsymbol{x}\right|D \right) d\boldsymbol{x} \approx \sum_{t = 1}^{T} h\!\left(\boldsymbol{x}^{\left(t\right)}\right) P\!\left(\boldsymbol{x}^{\left(t\right)}\right) = \sum_{t = 1}^{T}  h\!\left(\boldsymbol{x}^{\left(t\right)}\right) \frac{A^{\left(t\right)}}{Z^{\left(T\right)}} 
\end{equation}
It is \eqref{eq.6.32} that will us enable us to compute the first and second moments of the relevances defined on a Dirichlet distribution.

\section{The Issue of Confidence Bounds}
\label{chapter4.2}
Say we wish to assign the function $u$ to a bivariate probability distributions of the form: 

\begin{table}[h]
	\centering
		\begin{tabular}{c|c|c|c}
			 					& $b_{1}$							 				& $b_{2}$							    & 									 	       \\ \hline
	$a_{1}$				& $\theta_{1}$ 								& $\theta_{3}$ 				    & $\theta_{1}+\theta_{3}$	   \\ \hline
	$a_{2}$				& $\theta_{2}$ 								& $\theta_{4}$ 				    & $\theta_{2}+\theta_{4}$ 	 \\ \hline
								& $\theta_{1}+\theta_{2}$ 		& $\theta_{3}+\theta_{4}$ & $\theta_{1}+\theta_{2}+\theta_{3}+\theta_{4} = 1$ 				    			
		\end{tabular}
		\caption{Distribution 1}
		\label{tab.1}
\end{table}

\noindent Then, in practice, we only have indirectly access, by way of our data, to the probability distribution of interest; that is, rather then the known probabilities, Table~\ref{tab.1}, we only have some observed count data, Table~\ref{tab.2}:

\begin{table}[h]
	\centering
		\begin{tabular}{c|c|c|c}
			 					& $b_{1}$					& $b_{2}$						& 									 	\\ \hline
	$a_{1}$				& $r_{1}$ 				& $r_{3}$ 					& $r_{1}+r_{3}$	       \\ \hline
	$a_{2}$				& $r_{2}$ 				& $r_{4}$ 					& $r_{2}+r_{4}$ 	      \\ \hline
								& $r_{1}+r_{2}$ 	& $r_{3}+r_{4}$ 		& $r_{1}+r_{2}+r_{3}+r_{4} = n$ 				    			
		\end{tabular}
		\caption{Count data}
		\label{tab.2}
\end{table}

\noindent where $n$ is the total number of observations.

Let $D =\left(r_{1},r_{2},r_{3},r_{4}\right)$ be the observed count data in Table~\ref{tab.2}. Let $\boldsymbol{\theta} = \left(\theta_{1},\theta_{2},\theta_{3},\theta_{4}\right)$ be the vector of the unknown probabilities in Table~\ref{tab.1}. Then the likelihood function of the unknown probabilities is assumed to follow multinomial distribution: 
\begin{equation}
		\label{eq.5.3}
		L\!\left(\boldsymbol{\theta}\right) = p\!\left(\left.D\right|\boldsymbol{\theta}\right) 
		=\frac{\left(r_{1}+r_{2}+r_{3}+r_{4}\right)!}{r_{1}! r_{2}! r_{3}! r_{4}!}\theta_{1}^{r_{1}}\theta_{2}^{r_{2}}\theta_{3}^{r_{3}}\theta_{4}^{r_{4}}
\end{equation}
As a prior for the unknown probabilities we assign the uninformative Dirichlet prior 
\begin{equation}
	\label{eq.5.4}
	p\!\left(\boldsymbol{\theta}\right) \propto  \theta_{1}^{-1}\theta_{2}^{-1}\theta_{3}^{-1}\theta_{4}^{-1}
\end{equation}
which, if marginalized, collapses to the uninformative Beta prior. Combining the likelihood \eqref{eq.5.3} and prior\eqref{eq.5.4}, by way of the product rule, and normalizing, by way of the sum rule, we obtain the multivariate Dirichlet posterior distribution of the theta's given the observed count data:
\begin{equation}
	\label{eq.5.2}
	p\!\left(\left.\boldsymbol{\theta}\right|D\right) =  \frac{\left(r_{1}+r_{2}+r_{3}+r_{4} - 1 \right)!}{\left(r_{1}-1\right)! \left(r_{2}-1\right)! \left(r_{3}-1\right)! \left(r_{4}-1\right)!}\theta_{1}^{r_{1}-1}\theta_{2}^{r_{2}-1}\theta_{3}^{r_{3}-1}\theta_{4}^{r_{4}-1}
\end{equation}

Each realization of the function $u\!\left(\boldsymbol{\theta}\right)$ maps onto a corresponding probability $p\!\left(\left.\boldsymbol{\theta}\right|D\right)$. By arranging the values $u\!\left(\boldsymbol{\theta}\right)$ on the $x$-axis and the corresponding $p\!\left(\left.\boldsymbol{\theta}\right|D\right) d\boldsymbol{\theta}$ on the $y$-axis, we may obtain the univariate probability distribution of the function $u\!\left(\boldsymbol{\theta}\right)$. This probability distribution of $u\!\left(\boldsymbol{\theta}\right)$ takes the uncertainty into account we have in regards to the unknown $\boldsymbol{\theta}$ and, consequently, lets us put confidence bounds on this function. 

If we only have four unknown theta's we may use brute computational force to partition the domain of $\boldsymbol{\theta}$ and compute of for each partitioning the corresponding pair $\left[u\!\left(\boldsymbol{\theta}\right), p\!\left(\left.\boldsymbol{\theta}\right|D\right) d\boldsymbol{\theta}\right]$, after which we then order the $u\left(\boldsymbol{\theta}\right)$ and plot them together with their corresponding probabilities $p\left(\left.\boldsymbol{\theta}\right|D\right) d\boldsymbol{\theta}$. However, for large distributions, having many unknown $\boldsymbol{\theta}$'s, this quickly becomes unpractical because of the curse of dimensionality. 

In the next section we present an implementation of the Nested Sampling framework by Skilling that allows us to evaluate the probability distribution of $u\left(\boldsymbol{\theta}\right)$ for highly variate  distributions of $\boldsymbol{\theta}$, by way of Monte Carlo sampling. The implementation is accomplished by way of the Inner Nested Sampling algorithm. The Inner Nested Sampling algorithm allows us to sample uniformly from the constrained likelihood space of the $\boldsymbol{\theta}$'s, a necessary prerequisite of the Nested Sampling framework.

\section{Inner Nested Sampling}
Let $L$  be a likelihood function defined on a highly multivariate parameter space $\boldsymbol{\theta}$. Then Nested Sampling is a Monte Carlo framework with which this multivariate likelihood function $L\!\left(\boldsymbol{\theta}\right)$ may be evaluated. The Nested Sampling framework needs uniform samples within the multivariate geometry of some likelihood constraint  $L^{*}$ in order to work. However, this framework does not tell us how to obtain these samples, that is, its optimal implementation is an open-ended research question. In this paper we give an algorithm, called Inner Nested Sampling, that obtains such uniform samples.

The idea behind Inner Nested Sampling is that we obtain a set of differentials of the multivariate geometry of the initial likelihood constraint $L^{*}$ at iteration step $t = 0$ of Nested Sampling proper. These differentials are defined by a direction $\mathbf{e}$ and a radius $R\!\left(\mathbf{e}\right)$ and serve as a proxy for the actual geometry and have the nice property that they may be uniformly sampled. Furthermore as with each iteration step $t$ the geometry defined by likelihood constraint will shrinks, the radii $R\!\left(\mathbf{e}\right)$ may be updated so as to reflect this shrinkage. This then allows us to continue the uniform sampling of these differentials and, by proxy, the likelihood geometry of interest.  

\subsection{A change of variable}
Say, we have a $m$-variate parameter vector $\boldsymbol{\theta} = \left(\theta_{1},\ldots, \theta_{m}\right)$. Then the likelihood constraint $L\!\left(\boldsymbol{\theta}\right) \leq L^{*}$ defines some sub-domain $V^{*}$ of the total paramer space $V$ of $\boldsymbol{\theta}$. 

Now, if we have an (approximate) modus $\boldsymbol{\hat{\theta}}$ of the likelihood function $L\!\left(\boldsymbol{\theta}\right)$ we may translate the origin of $\boldsymbol{\theta}$ to the location of this modus and proceed to make a change of variables from a Cartesian coordinate system to unit-vector coordinates, \begin{equation}
	\label{eq.7.1}
		\boldsymbol{\theta} = \boldsymbol{\hat{\theta}}  + r \mathbf{e}
\end{equation}
where $\mathbf{e} = \left(e_{1},\ldots,e_{m}\right)$ is a point on the unit-sphere and $r$ is the distance from the $\boldsymbol{\hat{\theta}}$ to the likelihood constraint $L^{*}$ in the direction of $\mathbf{e}$.
Unit-vector coordinates map the parameter vector $\boldsymbol{\theta}$ to a radius $r$ and  $\left(m-1\right)$ non-redundant coordinates:
\begin{equation}
	\label{eq.7.2}
	\left(\theta_{1},\ldots, \theta_{k}\right) \mapsto \left(r,e_{1},\ldots, e_{m-1}\right)
\end{equation}
where the $m$th redundant coordinate of the unit vector $\mathbf{e}$ may be found through the identity
\[
e_{m} = 1 - \sum_{i=1}^{m-1} e_{i}^{2}
\] 	 
The differential of the unit-vector transformation is
\begin{equation}
	\label{eq.7.3}
	dV^{*} = \frac{R\!\left(\mathbf{e}\right)^{m}}{m} dS
\end{equation}
where $dS$ are equi-volume `patches' of the surface $S$ of the $m$-dimensional unit-sphere. Note that the surface $S$ itself has dimensionality $\left(m-1\right)$. The differentials \eqref{eq.7.3} are the volumes of pyramids with base $dS$ and height $R\!\left(e_{1}\cdots e_{m}\right)$. So, the integral
\begin{equation}
	\label{eq.7.4}
	V^{*} = \int dV^{*} = \int_{S} \frac{R\!\left(\mathbf{e}\right)^{m}}{m} dS
\end{equation}

\begin{figure}[!h]
	\centering
		\centerline{\includegraphics[width=0.50\textwidth]{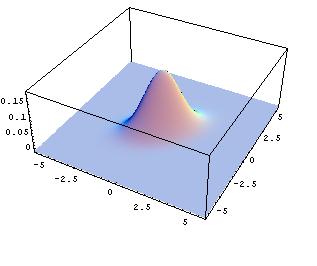}}
	\caption{Plot of Likelihood $L$}
	\label{Chap7a}
\end{figure}

\begin{figure}[!h]
	\centering
		\centerline{\includegraphics[width=0.40\textwidth]{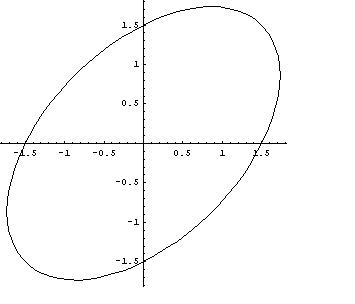}}
	\caption{Area Defined by Likelihood Constraint $L^{*} = 0.041$}
	\label{Chap7b}
\end{figure}

For example, say, we have the likelihood
\begin{equation}
	\label{eq.7.4q}
	L\left(x,y\right) = 0.184\exp\left[-\frac{1}{2}\left(x^{2} + x y + y^{2}\right)\right]
\end{equation}
which is given in Figure~\ref{Chap7a}. The likelihood constraint $L^{*} = 0.041$ corresponds with an ellips having `volume' $V^{*} = 8.162$, Figure~\ref{Chap7b}.

\begin{figure}[!h]
	\centering
		\centerline{\includegraphics[width=0.40\textwidth]{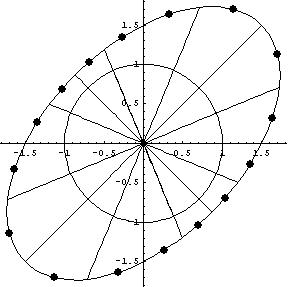}}
	\caption{Constrained Likelihood Space Partitioned by $dS$ into $dV^{*}$}
	\label{Chap7c}
\end{figure}

Since we have a two-dimensional likelihood space, the unit sphere is actually an unit circle and it follows that the surface $S$ is actually a circumference, where $S =2\pi $. The $dS$ is obtained by partitioning the circumference $S$ in $n$ equi-distant line elements, that is $dS = 2\pi/n$. Now, if we let $n = 16$, then
\begin{equation}
	\label{eq.7.4b}
	dS =\frac{\pi}{8}
\end{equation}
and the radii $R_{i}$, for $i=1,\ldots,n$ at the centers of these line elements $dS$ are shown in Figure~\ref{Chap7c}.

By summating the approximate differentials, \eqref{eq.7.3} and \eqref{eq.7.4b},
\begin{equation}
	\label{eq.7.5}
	dV^{*}_{i} = \frac{R_{i}^{m}}{m} dS = \frac{R_{i}^{m}}{m} \frac{\pi}{8}
\end{equation}
we may approximate the integral \eqref{eq.7.4}:
\begin{equation}
	\label{eq.7.6}
	\sum_{i=1}^{16} dV^{*}_{i} = 8.151 \approx 8.162 = V^{*} = \int dV^{*}  
\end{equation}
As we let $n\rightarrow \infty$, this approximation will become evermore accurate.

We summarize, the constrained likelihood space $V^{*}$ may be represented as a collection of differentials $\left\{dV^{*}_{1},\ldots,dV^{*}_{n}\right\}$. As it will turn out, it is trivially simple to sample uniformly from the set of the differentials $dV^{*}_{i}$, and by doing so we actually sample uniformly the space $V^{*}$ itself as we let $n\rightarrow \infty$, \eqref{eq.7.4}.

\subsection{Obtaining differentials}
The differentials depicted in Figure~\ref{Chap7c} were obtained through brute computational force. For highly multiariate likelihood spaces such methods are bound to fail due to the curse of dimensionality. However, Skilling's Nested Sampling framework comes here to the rescue, as we shall now demonstrate.

\begin{figure}[!h]
	\centering
		\centerline{\includegraphics[width=0.40\textwidth]{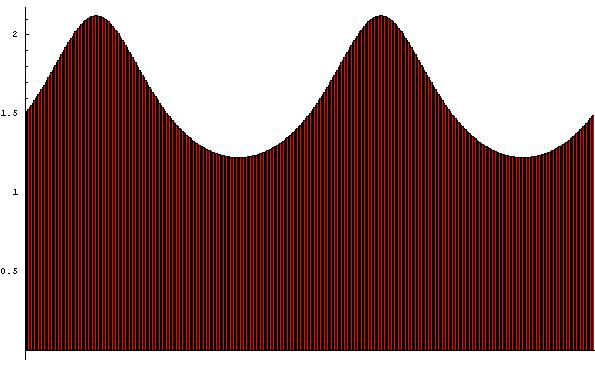}}
	\caption{Area Elements of $V^{*}$}
	\label{Chap7d}
\end{figure}

In analogy to the treatment of Nested Sampling proper in Chapter~6, we partion $dS$ in $n = 400$ equi-distant line elements, compute the area of the corresponding differentials $\left(dV_{1}^{*},\dots,dV_{400}^{*}\right)$, and plot them, Figure~\ref{Chap7d}.

\begin{figure}[!h]
	\centering
		\centerline{\includegraphics[width=0.40\textwidth]{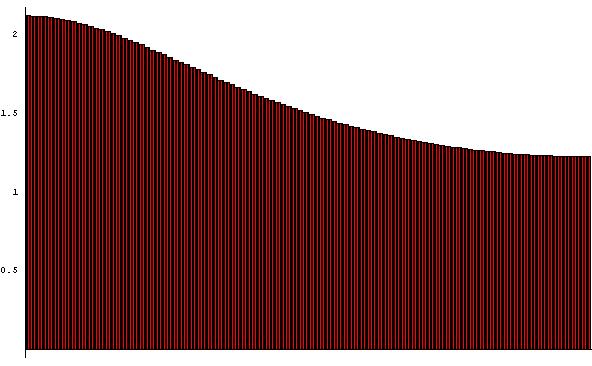}}
	\caption{Ordered Area Elements of $V^{*}$}
	\label{Chap7e}
\end{figure}

Again we are free to reorder these area elements as we like, Figure~\ref{Chap7e}.

\begin{figure}[!h]
	\centering
		\centerline{\includegraphics[width=0.40\textwidth]{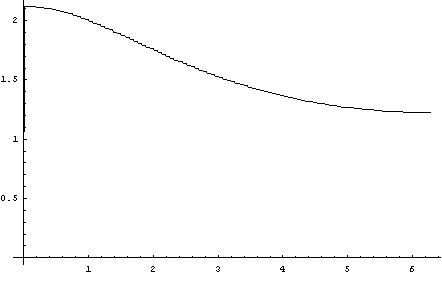}}
	\caption{Plot of Radii Function $g$}
	\label{Chap7f}
\end{figure}

Now all these rectangular elements have a base of  $dw = dS = 2\pi/400$. Being that there are 400 area elements we again might view Figure~\ref{Chap7f} as a representation of some monotonic descending function $g\left(w\right)$, where  $0\leq w\leq 2\pi$, Figure~\ref{Chap7f}.

Were we to integrate this function $g\left(w\right)$, we obtain an approximation of the area $V^{*}$, \eqref{eq.7.6}. However, this approximation is not our primary interest in Inner Nested Sampling. Though it may serve a check for those rare cases were we actually know the value of  $V^{*}$. Rather, the collection of differentials is what we are after, since these may sampled uniformly, thus, allowing us, by proxy, to uniformly sample $V^{*}$ itself.

The Nested Sampling framework, previously discussed, translates directly to Inner Nested Sampling, were we set $W$ to be the surface $S$ of the $m$-dimensional unit sphere
\begin{equation}
		\label{eq.7.7}
		W =  \frac{m \pi^{m/2}}{\Gamma\left(\frac{m}{2} + 1\right)}
\end{equation}
and take as the function to be evaluated
\begin{equation}
		\label{eq.7.8}
		f\!\left(\mathbf{e}\right) = \frac{R\!\left(\mathbf{e}\right)^{m}}{m} = g\!\left(w\right)
\end{equation}
In doing so, we end up with a collection of directions, $\mathbf{e}^{\left(q\right)}$, for $q=1,\ldots,Q$, with associated radii $R\!\left(\mathbf{e}_{q}\right)$, which are the distances from the modus $\boldsymbol{\hat{\theta}}$ to the constraint $L^{*}$, and weights $A^{\left(q\right)}$,  \eqref{Alg9}, \eqref{Alg10}, and \eqref{eq.7.8}:
\begin{align}
	\label{eq.7.9}
	A^{\left(q\right)} &= g^{\left(q\right)} dw^{\left(q\right)}  \nonumber\\
	\nonumber\\
	&=  g^{\left(q\right)} \left(w^{\left(q-1\right)} - w^{\left(q\right)}\right)\nonumber\\
	\\
	&=  \frac{R\!\left(\mathbf{e}^{\left(q\right)}\right)^{m}}{m} \left[\left(1-\frac{1}{N+1}\right)^{s-1}-\left(1-\frac{1}{N+1}\right)^{q}\right]  W  \nonumber
\end{align}
where $N$ is the number of objects used in the Inner Nested Sampling run and the $N$ directions $\mathbf{e}$ have been sampled under the constraint
\begin{equation}
	\label{eq.7.9a}
	R\!\left(\mathbf{e}_{n}^{\left(q\right)}\right) \leq \min R\!\left(\mathbf{e}_{r}^{\left(q-1\right)}\right)  \quad \quad n, r = 1, \dots, N
\end{equation}

The weights $A^{\left(q\right)}$ correspond with the volumes $dV_{q}^{*}$. By substituting \eqref{eq.7.7} in \eqref{eq.7.9}, we find, \eqref{eq.7.3}:
\begin{equation}
	\label{eq.7.9b}
	dV^{*}_{q} = A^{\left(q\right)}  = \frac{R\!\left(\mathbf{e}^{\left(q\right)}\right)^{m}}{m} dS_{q} 
\end{equation}
where
\begin{equation}
	\label{eq.7.9c}
	dS_{q} = \frac{1}{N+1} \left(1-\frac{1}{N+1}\right)^{q-1}  \frac{m\pi^{m/2}}{\Gamma\left(\frac{m}{2} + 1\right)}
\end{equation}
Note that for the differentials $dV^{*}$ and $dS$ we have let the subscripts ennumerate the iteration step.

\section{Obtaining uniform samples for Inner Nested Sampling}
The vector $\boldsymbol{\theta} = \left(\theta_{1},\dots,\theta_{m}\right)$ is a point in the parameter space and $L\left(\boldsymbol{\theta}\right)$ is the likelihood function that is defined on this parameter space. In highly dimensional problems the region of interest of $L\left(\boldsymbol{\theta}\right)$ becomes exponentially small relative to the total sample space. Therefore, the change that random sampling will yield a non-negligible likelihood value becomes exponentially small as the dimensionality of problem grows. 

In Inner Nested Sampling not the likelihood function $L\left(\boldsymbol{\theta}\right)$ is evaluated but the radius function $R\left(\mathbf{e}\right)$, where $\mathbf{e}$ is a point on the $m$-dimensional unit sphere. The radius function $R\left(\mathbf{e}\right)$ is more uniformly spread out over the surface of the unit-sphere than $L\left(\boldsymbol{\theta}\right)$ is over the parameter space. This makes random sampling of  the radius function much more feasible. Nonetheless, as the bound $R^{*}$ on $R\left(\mathbf{e}\right)$ becomes tighter and tighter, the number of rejections of the random draws will grow so much as to make a random walk necessary.   

\subsubsection{Random sampling of the unit sphere}
Random points on the unit-sphere are easily obtained by letting
\begin{equation}
	\label{eq.7.18}
	\mathbf{e} = \frac{\mathbf{e}_{0}}{\left\|\mathbf{e}_{0}\right\|},\quad \text{where}\ \ \mathbf{e}_{0} \backsim N\left(\mathbf{0},\mathbf{I}_{m}\right)
\end{equation}
A rejection occurs whenever $R\left(\mathbf{e}\right) < R^{*}$. If the number of rejections of the random draws of the $\mathbf{e}$ exceeds some number $M$, where $M >> 1$, a random walk on the unit sphere is performed. 

\subsubsection{Random walk on the unit sphere}
The random walk on the unit sphere takes at its starting point some $\mathbf{e}_{k}$ that has a radius $R_{k}$ which is known to satisfy the radius constraint $R_{k} > R^{*}$. and we set
\begin{equation}
	\label{eq.7.19a}
	\mathbf{e}^{\left(0\right)} = \mathbf{e}_{k}
\end{equation}
If the parameter space is 3-dimensional, then $\mathbf{e}^{\left(0\right)}$ will be a point on the surface of a 3-dimensional sphere. Then, for a given length $s$ of the random step to be taken, there is a 2-dimensional circle lying on the surface of this sphere with equi-probable possible candidates $\mathbf{e}^{\left(1\right)}$. To obtain a realisation $\mathbf{e}^{\left(1\right)}$, we sample a random point $\mathbf{u}$ on this circle, where
\begin{equation}
	\label{eq.7.19}
	\mathbf{v} = \frac{\mathbf{v}_{0}}{\left\|\mathbf{v}_{0}\right\|},\quad \text{where}\ \ \mathbf{v}_{0} \backsim N\left(\mathbf{0},\mathbf{I}_{2}\right)
\end{equation}
The circumference of the 2-dimensional circle whith the candidates $\mathbf{e}^{\left(1\right)}$ will change as the length $s$ of the random step changes. To be more more precise, let $\mathbf{e}^{\left(1\right)}$ be the proposal point of the first random step and let $\alpha$ be the angle between the starting point $\mathbf{e}^{\left(0\right)}$ and $\mathbf{e}^{\left(1\right)}$. Then the radius of the 2-dimensional circle will equal $\sin\alpha$, where $0 \leq \alpha \leq \pi$. 

If we translate the origin of our original coordinate system to $\left(\cos\alpha\right) \mathbf{e}^{\left(0\right)}$ and then rotate the axes of this 3-dimensional system so that the 2-dimensional circle, on which all the $\mathbf{e}^{\left(1\right)}$ are to be found, lies in the plane spanned by axes 2 and 3. Then the coordinates of $\mathbf{e}^{\left(1\right)}$ in this new coordinate system become, \eqref{eq.7.19}: 
\begin{equation}
	\label{eq.7.20}
	\tilde{\mathbf{e}}^{\left(1\right)} = \left(
\begin{array}{c}
	0 \cr \left(\sin\alpha\right) v_{1} \cr \left(\sin\alpha\right) v_{2} \cr
\end{array}
\right)
\end{equation}
In order to transform the coordinates of this alternative coordinate system, $\tilde{\mathbf{e}}^{\left(1\right)}$, back into the coordinates of the original coordinate system, $\tilde{\mathbf{e}}^{\left(1\right)}$, we must realize that in our original coordinate system the 2-dimensional circle of the $\tilde{\mathbf{e}}^{\left(1\right)}$ is orthogonal to $\left(\cos\alpha\right) \mathbf{e}^{\left(0\right)}$. This implies that the rotation matrix $B$, which accomplishes this re-transformation, has as its first column:
\begin{equation}
	\label{eq.7.21}
	\mathbf{b}_{1} = \mathbf{e}^{\left(0\right)} 
\end{equation}
The other two orthonormal columns of $B$, $\mathbf{b}_{2}$ and $\mathbf{b}_{3}$, then can be easily constructed by applying a Gram-Schmidt process. Let
\begin{equation}
	\label{eq.7.22}
	Q = \mathbf{I}_{3}
\end{equation}
Then
\begin{equation}
	\label{eq.7.23}
	\mathbf{b}_{2}^{\left(0\right)} = \mathbf{q}_{2} - \frac{\left\langle \mathbf{b}_{1},\mathbf{q}_{2}\right\rangle} {\left\langle \mathbf{q}_{2},\mathbf{q}_{2}\right\rangle} \mathbf{b}_{1} = \mathbf{q}_{2} - \left\langle \mathbf{b}_{1},\mathbf{q}_{2}\right\rangle \mathbf{b}_{1} 
\end{equation}
and
\begin{equation}
	\label{eq.7.24}
	\mathbf{b}_{2} = \frac{\mathbf{b}_{2}^{\left(0\right)}}{\left\|\mathbf{b}_{2}^{\left(0\right)}\right\|}
\end{equation}
Likewise, we have
\begin{equation}
	\label{eq.7.25}
	\mathbf{b}_{3}^{\left(0\right)} = \mathbf{q}_{3} - \left\langle \mathbf{b}_{1},\mathbf{q}_{3}\right\rangle \mathbf{b}_{1} - \left\langle \mathbf{b}_{2},\mathbf{q}_{3}\right\rangle \mathbf{b}_{2} 
\end{equation}
and
\begin{equation}
	\label{eq.7.26}
	\mathbf{b}_{3} = \frac{\mathbf{b}_{3}^{\left(0\right)}}{\left\|\mathbf{b}_{3}^{\left(0\right)}\right\|}
\end{equation}
This results in the rotation matrix
\begin{equation}
	\label{eq.7.27}
	B = \left(
\begin{array}{l l l}
	\mathbf{e}^{\left(1\right)} & \mathbf{b}_{2} & \mathbf{b}_{3}
\end{array}\right),\quad \text{where}\ \ B^{T}B = \mathbf{I}_{3}
\end{equation}
We then have that
\begin{equation}
	\label{eq.7.28}
	\mathbf{e}^{\left(1\right)} = \left(\cos\alpha\right) \mathbf{e}^{\left(0\right)} + B\tilde{\mathbf{e}}^{\left(1\right)}
\end{equation}
In the random walk the proposed step $\mathbf{e}^{\left(t\right)}$ is rejected whenever either, \eqref{eq.7.1}, 
\begin{equation}
\label{eq.7.29}
	L\left[\boldsymbol{\hat{\theta}} + R\left(\mathbf{e}^{\left(1\right)}\right) \mathbf{e}^{\left(1\right)}\right] < L^{*}
\end{equation}
or
\begin{equation}
\label{eq.7.30}
	R\left(\mathbf{e}^{\left(1\right)}\right) < R^{*}
\end{equation}

The random walk step must be modulated in order that the angle $\alpha$ becomes smaller as the number of rejections increases and, visa versa, larger as the number acceptances increases. Now, since a step length of $s$ corresponds with some angle $\alpha$, we are free to modulate $\alpha$ instead of $s$. So, let
\begin{equation}
	\label{eq.7.31}
	u^{\left(\tau\right)}\backsim U\left(0,1\right)
\end{equation}
where $\tau \geq t$ and $u^{\left(\tau\right)}$ should not equal 0 or 1. Let
\begin{equation}
	\label{eq.7.32}
	step^{\left(\tau\right)} = \left\{ 
	\begin{aligned}
		&step^{\left(\tau-1\right)} +1/2, \quad \text{if rejection} \\
		&step^{\left(\tau-1\right)} -1/2, \quad \text{if acceptance} 
	\end{aligned}
	  \right.
\end{equation}
where $step^{\left(0\right)} = 0$. Then, constraining $0 < \alpha^{\left(\tau\right)} < \frac{\pi}{2} $, we let
\begin{equation}
	\label{eq.7.33}
	\alpha^{\left(\tau\right)} =  \arcsin\left[ \left(u^{\left(\tau\right)}\right)^{step^{\left(\tau\right)}} \right] 
\end{equation}
where it is understood that $u^{\left(\tau\right)}$, \eqref{eq.7.29}, should not equal 0 or 1. 

For the case where the dimensionality of the parameter space is $m =3$, this procedure generalizes to
\begin{equation}
	\label{eq.7.34}
	\mathbf{e}^{\left(t\right)} = \left(\cos\alpha^{\left(\tau\right)}\right) \mathbf{e}^{\left(t-1\right)} + B^{\left(t-1\right)} \tilde{\mathbf{e}}^{\left(\tau\right)}
\end{equation}
where $t$ is the number of random steps accepted in the random walk, $\tau$ is the total number of iterations performed by the random walk, and, \eqref{eq.7.20},
\begin{equation}
	\label{eq.7.36}
	\tilde{\mathbf{e}}^{\left(\tau\right)} = \left(
\begin{array}{c}
	0 \cr \left(\sin\alpha^{\left(\tau\right)}\right) v_{1}^{\left(\tau\right)} \cr \left(\sin\alpha^{\left(\tau\right)}\right) v_{2}^{\left(\tau\right)} \cr
\end{array}\right)
\end{equation}
where, \eqref{eq.7.19},  
 \begin{equation}
	\label{eq.7.35}
	\left(
\begin{array}{c}
v_{1}^{\left(\tau\right)} \cr v_{2}^{\left(\tau\right)} \cr
\end{array}\right) = \frac{\mathbf{v}_{0}^{\left(\tau\right)}}{\left\|\mathbf{v}_{0}^{\left(\tau\right)}\right\|},\quad \ \ \mathbf{v}_{0}^{\left(\tau\right)} \backsim N\left(\mathbf{0},\mathbf{I}_{2}\right)
\end{equation}

In our preliminary empirical studies the number of random walk acceptances fluctuated around the $67\%$, which is as it should be. Note that this modulation of a random walk on the unit sphere is an adaptation of the modulating algorithm on the Cartesian parameter space as given by Skilling, \cite{Skilling04}.

\subsection{Drawing Uniform samples from Inner Nested Sampling realizations}
\label{section7.3}
If for some $m$-variate parameter space we have a collection of sampled unit-vectors $\left\{\mathbf{e}^{\left(1\right)},\ldots,\mathbf{e}^{\left(Q\right)}\right\}$ with corresponding radii $\left\{R^{\left(1\right)},\ldots,R^{\left(Q\right)}\right\}$. Then this allows us to draw a uniform sample from the corresponding differentials $\left\{dV_{1},\ldots,dV_{Q}\right\}$, and by doing so we actually sample uniformly the space $V^{*}$ itself as we let $Q\rightarrow \infty$, \eqref{eq.7.4}.
\\
\\
\noindent\textbf{Step 1}
\\
\noindent First we uniformly draw a differential $dV_{q}$. We do this by drawing from the uniform distribution:
\begin{equation}
	\label{eq.7.10}
	u \backsim U\left(0, Z^{\left(Q\right)}\right)
\end{equation}
where Z is understood to be the approximation of the $V^{*}$, \eqref{Alg11} and \eqref{eq.7.9}:
\begin{equation}
	\label{eq.7.11}
	Z = \sum_{q=1}^{Q} A^{\left(q\right)}
\end{equation}
We then find the smallest index value $q$ for which the following inequality holds, \eqref{eq.7.6}:
\begin{equation}
	\label{eq.7.12}
		\sum_{i=1}^{q} A^{\left(i\right)} \geq u
\end{equation}
This $q$ then is the index number of the uniformly drawn $dV_{q}$.
\\
\\
\noindent\textbf{Step 2}
\\
\noindent Let
\begin{equation}
	\label{eq.7.13}
	v = u - \sum_{i=1}^{q-1} A^{\left(i\right)}
\end{equation}
where $0\leq v \leq A^{\left(q\right)}$. Since the differentials $dV_{q}$ are pyramids in the limit $n\rightarrow\infty$, Figure~\ref{Chap7c}, the realisation $v$, $0\leq \nu \leq dV_{q}$, geometrically corresponds with the volume of a sub-pyramid having height $\rho$, where $0 \leq \rho \leq R_{q}$, that is
\begin{equation}
	\label{eq.7.14}
	v = \frac{\rho^{m}}{m} dS_{q} 
\end{equation}
where \eqref{eq.7.9c}:
\[
	dS_{q} = \frac{m \pi^{m/2}}{\Gamma\left(\frac{m}{2} + 1\right)} \frac{m}{m+1} \left(1-\frac{1}{m+1}\right)^{q-1} 
\]
From \eqref{eq.7.14}, we then have
\begin{equation}
	\label{eq.7.16}
	\rho = \sqrt[m]{\frac{m v}{dS_{q}}}
\end{equation}
\\
\noindent\textbf{Step 3}
The proposal of the uniformly sampled constrained likelihood space of $\boldsymbol{\theta}$, then simply becomes, \eqref{eq.7.1}:
\begin{equation}
	\label{eq.7.17}
	\boldsymbol{\theta}_{\text{proposal}} = \boldsymbol{\hat{\theta}}  + \rho \mathbf{e}^{\left(q\right)}
\end{equation}
where $\boldsymbol{\hat{\theta}}$ is the modus of the likelihood function which is evaluated, $\rho$ is \eqref{eq.7.16}, and $\mathbf{e}^{\left(q\right)}$ is the direction for which the radius $R^{\left(q\right)}$ was determined.

Integrals are limit cases of simple summation. In calculus differentials go to zero in order to obtain infinite precision. So, integrals in general have an infinite amount of differentials. Nested Sampling steps away from the limit case of calculus by approximating an infinite amount of differentials that go to zero with a finite amount of differentials greater than zero, which it then summates. Now, where we to sample to uniformly from $V^{*}$ then, because of identity \eqref{eq.7.4}, we may just as well sample uniformly over the infinite amount of differentials $dV^{*}$. 

Inner Nested Sampling, just like Nested Sampling, also steps away this limit case of infinite precision. By analogy, the finite amount of differentials that were obtained through random sampling are uniformly sampled, rather than the actual infinite amount of differentials. The finite set of differentials, obtained through random sampling, serves as a proxy for the infinite set of differential. So, by sampling uniformly over the finite set of differentials one approximately samples over $V^{*}$. The larger the set of differentials, the more accurate this approximation.

\subsection{Implementing Nested Sampling}
We now give the algorithm that integrates the Inner Nested Sampling within the Nested Sampling framework. 
\\
\\
\textbf{Inner Nested Sampling}
\\
Let the Inner Nested Sampling do a run for some large likelihood constraint that is found by shooting a large number of points in some preset parameter space of $\boldsymbol{\theta}$. The smallest likelihood value is taken as the initial likelihood constraint $L^{*}_{0}$. Let the Inner Nested Sampling algorithm explore the geometry defined by this $L^{*}_{0}$, using $M$ objects. Then we have an approximation of the corresponding domain $V^{*}$, in the form of the weights $A^{\left(q\right)}$, \eqref{eq.7.9b}, and the associated directions $\mathbf{e}^{\left(q\right)}$ and corresponding radii $R^{\left(q\right)} = R\!\left(\mathbf{e}^{\left(q\right)}\right)$. 
\\
\\

\noindent\textbf{Nested Sampling proper}
\\
Using the procedure in Section~\ref{section7.3}, draw $\left(N-1\right)$ random objects with likelihoods greater then $L^{*}_{0}$ from the collection of Inner Nested Sampling differentials $\left\{dV_{1}^{*}, \ldots, dV_{Q}^{*}\right\}$ and add these objects with the object corresponding with the initial likeklihood constraint $L^{*}_{0}$. We then have $N$ objects with which to perform Nested Sampling proper, as described in Chapter 6.

As the likelihood contour shrinks with every further Nested Sampling iteration, the probability grows that the proposed $\boldsymbol{\theta}$, \eqref{eq.7.17}, will be rejected. When the ratio of rejections becomes prohibitively large we update the monotonic function $g$ of the Inner Nested Sampling run, Figure~\ref{Chap7f}. This can be done by either determining the new $R^{\left(q\right)}$ corresponding with each stored direction vector $\mathbf{e}^{\left(q\right)}$. Or, by taking advantage of the fact that the likelihood contour generally will retain its shape as it shrinks, one alternatively may, if pressed for time, take with intervals values of $R^{\left(q\right)}$ on the $w$-`axis' and use some interpolating scheme for the other $R^{\left(q\right)}$ values. And sampling for Nested Sampling proper then continues for this updated function of  $g$ of the Inner Nested Sampling run. 

Note that the number of possible directions in the Nested Sampling algorithm is constrained to the number of vectors   stored in the Inner Nested Sampling run $\mathbf{e}$. In order to have many possible directions $\mathbf{e}$, a large number of objects may be taken be taken in the Inner Nested Sampling run. In our preliminary study we found that for a 20-variate normal likelihood $M = 1000$ initial Inner Nested Sampling objects resulted in $Q = 12.000$ stored direction vectors $\mathbf{e}$, from which $T = 2000$ random points were actually sampled in the Nested Sampling proper, which was initialized with $N = 100$ objects.    
	
\subsection{A  caveat}
Before we finish the outline of Inner Nested Sampling algorithm we have one more point to make. In Section~\ref{section7.3}, we sampled uniformly from the differential $dV_{q}^{*}$ by sampling the corresponding radius  $R^{\left(q\right)}$. This correspond geometrically with sampling the center line of the differential $dV_{q}^{*}$. Being that this center line itself is sampled uniformly from the unit sphere, either by way of randam draw or a random walk, this will still constitute a random draw in case this differential $dV_{q}^{*}$ is only sampled once during Nested Sampling proper. However, in the case were we draw the same differential twice, as we sample uniformly over the collection of differentials, uniformity is compromised, in that given a previous draw over that differential, we have pertinent knowledge of the possible realisations of the samples points. 

This may be remedied as follows. At the preliminary phase of Inner Nested Sampling algorithm run, as the radius constraint is not yet so strong that a random walk has to be employed, a very great amount of rejected directions $e$ are generated through pure random sampling. Now, instead of throwing these rejections away, we can store a predesignated number $q$ of the rejections which have radii closest to the radii $R^{\left(q\right)}$ already in place in a `safety matrix'. This safety matrix is updated as the number of rejections, thus, getting radii which are evermore closer the $R^{\left(q\right)}$ already in place. 

Now, it is expected that this procedure will go a long way to remedy the possible distortion of uniformity in the sampling within the differentials themselves, since it is especially these differentials $dV_{q}^{*}$ in the preliminary phase of Inner Nested Sampling which are most likely to be sampled more than once. As can be deduced from \eqref{Alg11b}:
\begin{equation}
\label{eq.7.37}
	dV_{q}^{*} = A^{\left(q\right)}  \rightarrow 0
\end{equation}
as $q \rightarrow \infty$. 
However, should it happen that one of the differentials $dV_{q}^{*}$ obtained through a random walk is sampled more than once, however unlikely, two choices are left to obtain extra samples for this differentials. Either try to obtain a $\tilde{R}^{\left(q\right)}$ by performing a random walk under the contraint
\begin{equation}
\label{eq.7.38}
	 \frac{1}{2} R^{\left(q-1\right)} \leq \tilde{R}^{\left(q\right)} \leq \frac{1}{2} R^{\left(q+1\right)}
\end{equation}
whenever in the uniform sampling of the differentials a $R^{\left(q\right)}$ is drawn which already has been drawn once. Or, alternatively, ignore possible distortion in uniformity, especially since the $dV_{q}^{*}$ more and more become like true differentials, that is, infinitely thin slices of volume, as $q \rightarrow \infty$, \eqref{eq.7.37}. And the ever mounting computational cost in exploring the radius constraint \eqref{eq.7.38} will eventually not outweigh the added benefits that the object $\tilde{R}^{\left(q\right)}$ has to offer in terms of the restoration of uniformity, since the distortions in uniformity are so minute.

\section{Resolving the Issue of Confidence Bounds}
Let $\theta_{ij} = \left(\theta_{11},\theta_{12},\ldots,\theta_{IJ}\right)$ for $i = 1, \dots, I$, $j = 1, \dots, J$, where $I \times J = M$. Also let $\theta_{i+} = \left(\theta_{1+},\ldots,\theta_{I+}\right)$ and $\theta_{+j} = \left(\theta_{+1},\ldots,\theta_{+J}\right)$. The issue of interest were will apply the Inner Nested Sampling is to evaluate the value $u\left(\theta_{ij}\right)$. This evaluation is over all plausible values of the $\theta_{ij}$, weighted according to the multivariate posterior distribution of these theta's, \eqref{eq.5.2}: 
\begin{equation}
	\label{eq.8.2}
	p\left(\left.\theta_{ij}\right|D\right) \propto  \theta_{11}^{r_{11}-1}\theta_{12}^{r_{12}-1}\cdots \theta_{IJ}^{r_{IJ}-1}
\end{equation}
where $D=\left(r_{11},r_{12},\dots,r_{IJ}\right)$.

In the case of a large $M$ the brute computational evaluation of the integrals which give us, respectively, the first and second momemt of the function $u\left(\theta_{ij}\right)$:
\begin{equation}
	\label{eq.8.3}
	M_{1} = \int  p\left(\left.\theta_{ij}\right|D\right) u\left(\theta_{ij}\right) d\theta_{ij}
\end{equation}
and
\begin{equation}
	\label{eq.8.3b}
	M_{2} = \int  p\left(\left.\theta_{ij}\right|D\right) \left[u\left(\theta_{ij}\right)\right]^{2} d\theta_{ij}
\end{equation}
quickly becomes unpractical because of the curse of dimensionality. So, the Nested Sampling framework, as implemented by the Inner Nested Sampling algorithm, will be used to evaluate \eqref{eq.8.3}. This is done by evaluating, respectively, the sums, \eqref{eq.7.36}:
\begin{equation}
	\label{eq.8.4}
	M_{1} \approx \sum_{t=1}^{T}  p\left(\boldsymbol{\theta}_{t}\right) u\left(\boldsymbol{\theta}_{t}\right) = \sum_{t=1}^{T} \frac{A_{t}}{Z^{\left(T\right)}} u\left(\boldsymbol{\theta}_{t}\right)
\end{equation}
and
\begin{equation}
	\label{eq.8.4b}
	M_{2} \approx \sum_{t=1}^{T}  p\left(\boldsymbol{\theta}_{t}\right) \left[u\left(\boldsymbol{\theta}_{t}\right)\right]^{2} = \sum_{t=1}^{T} \frac{A_{t}}{Z^{\left(T\right)}} \left[u\left(\boldsymbol{\theta}_{t}\right)\right]^{2}
\end{equation}
By way \eqref{eq.8.4} and \eqref{eq.8.4b}, we may put the following approximate confidence bounds on the relevance $u$:
\begin{equation}
	\label{eq.8.4c}
	M_{1} - \sqrt{M_{2}-M_{1}^{2}} \leq u \leq M_{1} + \sqrt{M_{2}-M_{1}^{2}}
\end{equation}

\subsection{Inner Nested Sampling for the Dirichlet Distribution}
As it turns out, the dirichlet distribution, that is, the multinomial beta distribution, \eqref{eq.8.2} is particullary amenable to the Inner Nested Sampling algorithm, since it has known modus as well as analytical derivable $R\!\left(\mathbf{e}\right)$ for the first run, when no constraints radius constraint is yet in place, other than the simplex form of the domain on which \eqref{eq.8.2} is defined. This is particullary helpful, since in general we have that the finding of the radii $R\!\left(\mathbf{e}\right)$ is the most computational intensive part of the Inner Nested Sampling, as it requires some kind of search algorithm, such as slice sampling, for example. 

We now revert back to the notation of Section~\ref{chapter4.2}, and let
\begin{equation}
	\label{eq.8.4d}
   \boldsymbol{\theta} = \theta_{jk} = \theta_{i} = \left(\theta_{1},\ldots,\theta_{M}\right)
\end{equation}
and
\begin{equation}
	\label{eq.8.4e}
	p\left(\left.\boldsymbol{\theta} \right|D\right) \propto  \theta_{1}^{r_{1}-1}\cdots \theta_{M}^{r_{M}-1}
\end{equation}
The modus of \eqref{eq.8.4e}, using some mathematical package, may be derived as:
\begin{equation}
	\label{eq.8.5}
		\hat{\boldsymbol{\theta}} = \hat{\theta}_{i} =\left(\frac{r_{1}}{n}, \ldots, \frac{r_{M}}{n}\right)
\end{equation}
where $n =  \sum_{i} r_{i}$. The radius constraint, for any modus $\hat{\boldsymbol{\theta}}$, is the distance from this modus to the boundary of the $\left(M-1\right)$-simplex defined by the vertices
\begin{align}
	\label{eq.8.6}
		\mathbf{v}_{1} &= \left(1, 0, \ldots, 0\right)  \nonumber \\
		\mathbf{v}_{2} &= \left(0, 1, \ldots, 0\right)  \nonumber \\
											 & \vdots \\
		\mathbf{v}_{M} &= \left(0, 0, \ldots, 1\right)  \nonumber
\end{align}
where the $\left(M-1\right)$-simplex is the $M$-dimensional domain of the Dirichlet parameters $\boldsymbol{\theta}$.

The $\left(M-1\right)$-simplex is spanned by the $\left(M-1\right)$ vectors, \eqref{eq.8.6},
\begin{align}
	\label{eq.8.7}
		\mathbf{u}_{1} &= \mathbf{v}_{1} - \mathbf{v}_{M} = \left(1, 0, \ldots, 0, -1\right)  \nonumber \\
		\mathbf{u}_{2} &= \mathbf{v}_{2} - \mathbf{v}_{M} = \left(0, 1, \ldots, 0, -1\right)  \nonumber \\
											 & \vdots \\
		\mathbf{u}_{M-1} &= \mathbf{v}_{M-1} - \mathbf{v}_{M} = \left(0, 0, \ldots, 1, -1\right)  \nonumber
\end{align}
The coordinate system defined by \eqref{eq.8.7} is not orthonormal. And in order to make it so, we first orthogonalize the system by way of a Gramm-Schmidt process. First, arbitrily, let 
\begin{equation}
	\label{eq.8.8}
	\mathbf{b}_{1} = \mathbf{u}_{1}
\end{equation}
Then 
\begin{equation}
	\label{eq.8.9}
	\mathbf{b}_{2} = \mathbf{u}_{2} - \frac{\left\langle \mathbf{b}_{1}, \mathbf{u}_{2} \right\rangle}{\left\langle \mathbf{u}_{2}, \mathbf{u}_{2} \right\rangle} \mathbf{b}_{1}
\end{equation}
and
\begin{equation}
	\label{eq.8.10}
	\mathbf{b}_{3} = \mathbf{u}_{3} - \frac{\left\langle \mathbf{b}_{1}, \mathbf{u}_{3} \right\rangle}{\left\langle \mathbf{u}_{3}, \mathbf{u}_{3} \right\rangle} \mathbf{b}_{1} - \frac{\left\langle \mathbf{b}_{2}, \mathbf{u}_{3} \right\rangle}{\left\langle \mathbf{u}_{3}, \mathbf{u}_{3} \right\rangle} \mathbf{b}_{2}
\end{equation}
etc., until $\mathbf{b}_{M-1}$. Then, normalizing the orthogonal vectors \eqref{eq.8.10}, 
\begin{equation}
	\label{eq.8.11}
		\mathbf{w}_{1} = \frac{\mathbf{b}_{1}}{\left\|\mathbf{b}_{1}\right\|}	\quad  \cdots \quad \mathbf{w}_{M-1} = \frac{\mathbf{b}_{M-1}}{\left\|\mathbf{b}_{M-1}\right\|}  
\end{equation}
we obtain the $\left(M-1\right)$-dimensional orthonormal coordinate system $W$ we are looking for:
\begin{equation}
	\label{eq.8.12}
	\mathbf{W} = \left[\begin{array}{l l l}
									\mathbf{w}_{1} & \cdots & \mathbf{W}_{M-1}
							 \end{array} \right]
\end{equation}

Let $\mathbf{e}$ be a random point on the $M-1$-dimensional unit-sphere, obtained by letting, \eqref{eq.7.18},
\begin{equation}
	\label{eq.8.13}
	\mathbf{e} = \frac{\mathbf{e}_{0}}{\left\|\mathbf{e}_{0}\right\|},\quad \text{where}\ \ \mathbf{e}_{0} \backsim N\left(\mathbf{0},\mathbf{I}_{M-1}\right)
\end{equation}
And let 
\begin{equation}
	\label{eq.8.14}
 \boldsymbol{\delta} = \mathbf{W} \: \mathbf{e}
\end{equation}
Then any point 
\begin{equation}
	\label{eq.8.15}
				\mathbf{p} = \hat{\boldsymbol{\theta}} + R \: \boldsymbol{\delta},		\quad R \geq 0
\end{equation}
will lie on the same `plane' as the $\left(M-1\right)$-simplex. In order to get the maximum radius $R$ for which $\mathbf{p}$ is a point on the the bondary of the $\left(M-1\right)$-simplex, we must realize that on this bondary one of the elements of the $\mathbf{p}$ must necessarily be zero. So we solve for every element of $\mathbf{p}$:     
\begin{equation}
	\label{eq.8.16}
				p_{i} = \hat{\boldsymbol{\theta}}_{i} + R_{i} \boldsymbol{\delta}_{i} = 0
\end{equation}
and obtain	
\begin{equation}
	\label{eq.8.17}
				R_{i} =  - \frac{\hat{\boldsymbol{\theta}}_{i}}{\boldsymbol{\delta}_{i}} 
\end{equation}
As only positive radii $R_{i}$ are allowed, the maximum radius $R$ for which $\mathbf{p}$ still lies on the boundary of the $\left(M-1\right)$-simplex is
\begin{equation}
	\label{eq.8.18}
				R = \min R_{i},	\quad \text{for those } R_{i} \geq 0 
\end{equation}
As any radius greater than $R$ will automtically result in negative $p_{i}$ values, which then would imply that $\mathbf{p}$ lies outside of the $\left(M-1\right)$-simplex.  

We summarize, in order to use Inner Nested Sampling on the $M$-dimensional Dirichlet distribution \eqref{eq.8.4e}, 
we draw a random point on the $\left(M-1\right)$-dimensional unit-sphere, \eqref{eq.8.13}, which we then transform to the direction \eqref{eq.8.14}. And the corresponding distance $R$ from the modus $\hat{\boldsymbol{\theta}}$ to the boundary of the domain of the Dirichlet parameters, $\boldsymbol{\theta}$, which is the $\left(M-1\right)$-simplex defined by the vertices \eqref{eq.8.6}, is found by way of \eqref{eq.8.17} and \eqref{eq.8.18}. Furthermore, the volume of this $\left(M-1\right)$-simplex is known to be
\begin{equation}
	\label{eq.8.19}
	V^{*} = \frac{\sqrt{M}}{\left(M-1\right)!}
\end{equation}
and \eqref{eq.8.19} then may serve as check on the accuracy of the estimation, \eqref{eq.7.36},
\begin{equation}
	\label{eq.8.20}
	\log Z^{\left(T\right)} \approx \log V^{*} \approx \left(M-1\right) \log\left(M-1\right) - \left(M-1\right) - \frac{1}{2} \log M 
\end{equation}
\\
\\



\begin{thebibliography}{9}



\bibitem {Skilling04}
Skilling, J.: Nested Sampling, In Bayesian Inference and Maximum Entropy Methods in Science and Engineering, (eds. Erickson G., Rychert J.T., and Smith C.R.) AIP Conference Proceedings, American Institute of Physics, New-York. 395-405, (2004).

\bibitem {Skilling06}
Skilling, J.: Nested Sampling for General Bayesian Computation, Bayesian Analysis 4, pp. 833-860, (2006). 

\bibitem {Skilling08}
Skilling, J.: The Canvas of Rationality, In Bayesian Inference and Maximum Entropy Methods in Science and Engineering, Sao Paulo, Brazil 2008 (eds. Lauretto MS, Pereira CAB) AIP Conference Proceedings, American Institute of Physics, New-York. 67-79, (2008). 

\bibitem {Zellner71}
Zellner A.: An Introduction to Bayesian Inference in Econometrics, J. Wiley \& Sons, Inc., New York, (1971). 

\bibitem {MacKay03}
MacKay D.J.C.: Information Theory, Inference, and Learning Algorithms, Cambridge University Press, Cambridge, (2003).

\end{thebibliography}
\end{document}